\documentclass[runningheads]{llncs}
\usepackage{graphicx}
\usepackage{cite}
\usepackage{nameref}
\usepackage{amsmath,amsfonts,amssymb}
\usepackage{bm}
\usepackage{float}
\begin{document}
\title{The Duet of Representations and How Explanations Exacerbate It}
%
%
\author{Charles Wan \inst{1} \orcidID{0000-0001-8284-1353} \and Rodrigo Belo \inst{2} \orcidID{0000-0001-6839-5086} \and 
Leid Zejnilović \inst{2} \orcidID{0000-0002-4209-4637} \and Susana Lavado \inst{2} \orcidID{0000-0002-1088-6357}}
\authorrunning{Wan et al.}
%
\institute{Rotterdam School of Management, Erasmus University  \\ \email{wan@rsm.nl} \and 
Nova School of Business and Economics, Universidade NOVA de Lisboa \\ \email{\{rodrigo.belo,leid.zejnilovic,susana.lavado\}@novasbe.pt}}

\maketitle              
\begin{abstract}
An algorithm effects a causal representation of relations between features and labels in the human's perception. Such a representation might conflict with the human's prior belief.  Explanations can direct the human's attention to the conflicting feature and away from other relevant features. This leads to causal overattribution and may adversely affect the human's information processing.  In a field experiment we implemented an XGBoost-trained model as a decision-making aid for counselors at a public employment service to predict candidates' risk of long-term unemployment.  The treatment group of counselors was also provided with SHAP.  The results show that the quality of the human's decision-making is worse when a feature on which the human holds a conflicting prior belief is displayed as part of the explanation.

\end{abstract}
\section{Introduction}

Artificial intelligence is increasingly embedded in everyday business and consumer decision-making.  For an array of reasons --- technical, psychological, organizational, legal, and ethical --- these decision-making systems are seldom fully automated.  They require human input or, as components of larger socio-technical systems, interface to a significant degree with other components that are predominantly human-driven.  Understanding how humans interact with algorithms \textit{epistemically}, therefore, is of crucial importance whether the considerations are primarily economic and managerial or societal and ethical.  

One principal way algorithms communicate with humans is via causal representations \cite{wan2022explainability}.  The algorithm effects a causal representation relating features and labels \cite{delanda2021materialist} in the human's perception.  This is conveyed through the algorithm's observable output --- typically predictions.  For example, imagine a context where humans are using an algorithm as a decision-making aid to predict the risk of loan default for different individuals.  Upon observing the algorithm's output (predictions), the human might attribute to the algorithm a causal representation with the following semantics: longer job tenure leads to a lower risk of loan default.

Informed by history, humans form prior beliefs with respect to predictive tasks \cite{lu2008bayesian}.  For example, in predicting loan default humans might associate lower income with a higher risk of loan default.  This representation may conflict with the causal representation attributed to the algorithm.  We call such a conflict the duet of representations. Because humans value simplicity, both their prior beliefs and the causal representations they attribute to algorithms are likely to be linear and sparse \cite{lombrozo2007simplicity}.  

One way to facilitate communication between agents is through the provision of explanations.  Explanatory methods can be viewed as tools that render the algorithm's causal representations human-interpretable.  They extract simple representations via some model additive to the original algorithm.  For example, \textsf{LIME} \cite{ribeiro2016should} use locally linear models to extract linear representations of relations between features and labels.  \textsf{SHAP} \cite{lundberg2017unified} use cooperative games with features as players and extract representations in the form of sets of important features.  Explanations direct human \textit{attention} to sparse and cogent representations with clear semantics.  They therefore increase the salience of any conflict with human priors.  We present empirical evidences from a field experiment that explanations exacerbate the duet of representations and affect the quality of human decisions.

The results of our study suggest that fruitful and robust human-algorithm interaction requires a reconsideration of what constitutes ``communication'' between the algorithm and the human.  Epistemically and communicatively, it is not sufficient to extract causal representations effected by algorithms.  Effective communication depends on understanding the whys of a causal representation effected by the algorithm --- the standpoint from which it is generated --- as well as reciprocity, negotiability and the ability to refer to a shared objective reality.  A set of desiderata for human-algorithm interaction is offered in Section \ref{discussion}.          

\section{Theory and Related Work}

\subsection{Human-algorithm Collaboration}

One literature stream on human-algorithm interaction adopts a managerial or engineering perspective and takes performance as the object of study.  What matters is not the epistemic content of human-algorithm interaction but the effect it has on performance, evaluated against some metric.  In this vein, \cite{fugener2021will} analyze how human-algorithm interaction can lead to the ``cyborgization'' of human thought. This results in the loss of unique human knowledge that can contribute to effective decision-making.  \cite{fugener2022cognitive} show that humans and AI working together can outperform AI alone when the latter delegates to the former.  \cite{sun2022predicting} study human deviations from algorithmic prescriptions in warehouse operations.  They devise a machine learning algorithm to predict the deviations and show that incorporating them in logistics planning improves performance.  Results from \cite{kawaguchi2021will}'s field experiment show that the failure to adopt algorithmic recommendation can lead to a gap between the nominal and the actual performance of the algorithm.  \cite{gao2021human} develop a method that uses learning from bandit feedback to optimize human-AI collaboration.

The common thread behind this body of work is the conception of human-algorithm collaboration as a process that can and should be optimized in order to improve performance.  To the extent that the process produces observable output in experiments, simulations, and real-world deployment, the data are used to understand decision quality and how it can be enhanced.  This is an intellectual viewpoint that focuses on the ``external'' --- measurements whose variances are related to observed variances in the structure of human-algorithm interaction.  What is relatively less theorized is the epistemic content of this interaction.  That is, beyond the fact that human knowledge and human actions can affect performance, how do humans process algorithmic predictions or recommendations \textit{as information}? 

\subsection{Affective States}

Another stream of literature focuses on internal psychological states induced by interaction with the algorithm.  While the implicit goal might still be the improvement of a performance measure such as adoption, this body of work seeks to explain engagement psychologically or develop a normative framework for judging the conditions under which an internal psychological state is desirable.  The literature has identified chiefly two affective states as important for human-algorithm interaction --- trust and aversion.  \cite{dietvorst2015algorithm} introduces the concept of ``algorithm aversion'' and \cite{dietvorst2018overcoming} shows that agency helps to attenuate it.  \cite{lebovitz2022engage} investigates how in medical diagnosis the opacity of AI diagnostics can lead to a loss of trust via an increase in epistemic uncertainty.  \cite{jacovi2021formalizing} formalizes the notions of trust and trustworthiness in human-AI interaction and examine the criteria under which trust is normatively warranted.  \cite{glikson2020human} reviews tangibility, transparency, reliability, and immediacy as factors that help to inculcate cognitive trust and anthropomorphism as a factor that helps to inculcate emotional trust in the AI.  Lastly, \cite{ullman2018does} develops a multidimensional measure of trust in the context of robotics.  

What underlies this body of work is an ``internal'' view that psychological states are determined by how humans interface with the algorithm and in turn drive aspects of human-algorithm interaction.  Furthermore, there is a normative dimension in so far as certain psychological states such as trust are only warranted under specific conditions. 

\subsection{The Duet of Representations} \label{2.3}

While an affective state clearly disposes the human towards a particular set of actions vis-à-vis the algorithm, it lacks the adaptive rationality that allows agents to respond to changes.  Feelings of trust and aversion, once developed, are relatively constant, at least on the timescale of human-algorithm interaction \cite{dietvorst2015algorithm,dietvorst2018overcoming,lebovitz2022engage}.  They do not explain individual instances of interaction.  Consider interpersonal dynamics.  One might be inclined to accept suggestions from a trustworthy friend, with such inclination assumed to be relatively stable over time.  There will nonetheless be variability in one's actions if one is not to surrender one's agency completely.  If agency is the ability to perform a difference-making action (e.g. accept or reject a suggestion) as it pertains to one's goal, then affective states alone cannot account for the differences in actions.    
What enables difference-making actions and agency is representation.  Humans build mental models of the world and also attribute mental models --- via theory of mind --- to other agents \cite{lagnado_2021}. 
These models are often causal representations with interventionist\footnote{Setting $X$ to $x_0$, $Y$ would be $y_0$} or counterfactual\footnote{Had $X$ been $x_0$, $Y$ would have been $y_0$} semantics \cite{pearl2018book,woodward2021causation}.  Human collaboration requires, in addition to a representation of the shared goal or task, representations of other agents --- more precisely, representations of other agents' representations of the task \cite{xiang2022collaborative}.  When the human and the algorithm cooperate on a predictive task, it is the human's representation and the representation that she attributes to the algorithm that jointly enable the human to exercise her agency and perform difference-making actions. 

Formally, the human constructs a representation $R_{h}$ with respect to the predictive task from the space of human-interpretable representations $\mathcal{R}$.  This representation could be a causal model with interventionist or counterfactual semantics \cite{pearl2018book,woodward2021causation}.  It could also be a simpler heuristic \cite{gigerenzer1996reasoning,Gigerenzer:1999p256}.  For example, the representation for a binary classification task might be a sparse set of feature values that contributes to a negative prediction: $\{U=u_0\} \lor \{V=v_0\} \lor \{W=w_0\} \mapsto -1$.  Since this representation defines the human's state of knowledge before using the algorithm, it can also be understood as the prior belief.  The algorithm also effects a causal representation that relates features and labels.  For example, a causal representation (that the human attributes to the algorithm) for a binary classification task might be that certain features values cause a positive prediction whereas others cause a negative prediction: $\{U=u_0\} \lor \{V=v_0\} \lor \{W=w_0\} \mapsto 1; \{X=x_1\} \lor \{Y=y_1\} \lor \{Z=z_1\} \mapsto -1$.  

Given an instance from the input space $\mathcal{I}$, the representation the algorithm effects and the human prior belief interact to induce a human action from the action space $\mathcal{A}$: $\mathcal{I} \times \mathcal{R} \times \mathcal{R} \mapsto \mathcal{A}$.  Using the binary classification example from above, consider the input instance $\{U=u_0\}$ for which the algorithm gives a positive prediction.  The causal representation attributed to the algorithm $R_{a}: \{U=u_0\} \mapsto 1$ conflicts with the prior belief $R_{h}: \{U=u_0\} \mapsto -1$.  This might prompt the human to reject or revise the algorithm's prediction.  The set of possible actions as well as the exact process by which an action is selected will vary by the predictive task.  Bayesian updating \cite{bundorf2019humans}, for example, might be appropriate for continuous labels.    

\subsection{Explanations as Compressed Representations}

\cite{simon2013administrative} defines communication as ``any process whereby decisional premises are transmitted from one member of an organization to another.''  Effective collaboration requires that the agents' states of knowledge be commensurate with the decision-making process.  Sometimes routines and procedures encode past learning and constrain the agents to cooperate \cite{cyert1963behavioral}.  At other times explicit communication between the agents is needed to establish an adequate basis for action.  With respect to predictive tasks where the human shares in decisional authority and responsibility, the latter means that the algorithm's representation would have to be communicated to the human.  Explanations can render complex representations human-interpretable and help to close the gap in shared knowledge of decisional premises.  Counterfactual and causal explanations \cite{galhotra2021explaining,schwab2019cxplain} especially might be more aligned with human mental models. 

In this capacity, explanations can be regarded as compressed representations: $R_{a} = \psi(h)$, where $|R_{a}| < |h|$.  That is, an explanatory method $\psi(\cdot)$ extracts a compressed representation $R_{a}$ of the original function $h$ learned by the algorithm from the space of human-interpretable representations $\mathcal{R}$.  The compressed representation $R_{a}$ is more sparse than the underlying function $h$ in some sense, e.g. the number of features.  For example, \textsf{LIME} approximate the underlying model locally with sparse linear representations \cite{ribeiro2016should}.  \textsf{SHAP} model features as players in a cooperative game and extract the most relevant ones as explanations \cite{lundberg2017unified}.  Both have human-interpretable semantics --- the former in the form of a sparse linear model, the latter in the form of a sparse set of relevant features.  

A number of studies \cite{guidotti2021evaluating,nauta2022anecdotal,han2022explanation,slack2021reliable} have developed frameworks for evaluating the fidelity of explanatory methods with respect to the original models.  Compressed representations generated by explanatory methods might also conflict with human priors.  Explanations increase the \textit{salience} of any conflict with human priors by commanding cognitive attention and directing it to sparse and cogent representations with clear semantics.  The fact that an explanation is explicitly and concisely shown to the human --- for \textsf{SHAP} it would be a set of relevant features --- can make the disagreement more conspicuous.  This exacerbates the conflict and can affect the quality of human decision-making.    

\cite{colin2022cannot} takes the view that explanations help human users to learn to meta-predict model predictions.  While achieving an understanding of the model sufficient for meta-predicting its predictions is a normative good that can be useful for many tasks, this conceptualization underplays the fact that in many real-world settings learning per se is not the express goal --- action is.  It cannot be assumed that a human user would suspend her prior beliefs in making decisions as she might when the objective is explicitly learning.  An explanation, therefore, is not simply a piece of information to be used for improving one's understanding.  It represents a distinct epistemic standpoint which can conflict with that of the human user to an extent that is consequential for actions.

\subsection{Psychological Salience and Information Processing}      

An explanation directs attention to a sparse representation with clear semantics.  This increases the \textit{salience} of features that form part of the explanation, especially if the human user has a strong prior belief on how they should be related to the label and this prior belief conflicts with what is implied by the explanation.  For example, the human might hold the belief that $\{U=u_0\} \mapsto -1$ and an explanation that $\{U=u_0\}$ contributes to a positive prediction would put the conflict in the crosshairs.       

Research in psychology has shown that salience can have a large impact on human judgment \cite{taylor1978salience, taylor1979generalizability} and even affect causal attribution \cite{taylor1975point, fiske1982structural}.  In the context of machine learning explanations, the focus of attention on a sparse set of features for which the human user holds strong prior beliefs can induce overconfidence \cite{wan2022explainability}.  By directing attention to features on which the human user has a conflicting prior belief, the explanation also directs attention \textit{away} from other features which could have been part of the human user's information processing.  This leads to causal overattribution.  The conflict also exacts a cognitive cost.  Both can affect the quality of human decision-making.        

\section{Field Experiment}

\subsection{Setting}

The empirical context is a public employment service (PES) in the European Union.  The PES provides services such as job referral and vocational training to unemployed individuals.  When an individual becomes unemployed, she has to register at PES to receive financial support from the government. During the registration process, usually done in person, the registrant gives her data to a counselor who will review her case and support her in finding employment.  According to an internal regulation, a counselor at the PES is obliged to assess the unemployed candidate's risk of \textsf{LTU} (long-term unemployment) upon registration, where \textsf{LTU} is defined as being involuntarily unemployed for a year or more.  

We trained and implemented an \textsf{XGBoost} classification model that took as input candidate features and returned as output a raw probability score (risk score) and a risk assessment.  Raw probability scores of \textsf{LTU} produced by \textsf{XGBoost} were converted into risk assessments of \textsf{low}, \textsf{medium}, and \textsf{high}, where \textsf{high} means a high probability of \textsf{LTU}. A risk assessment of \textsf{high} is equivalent to a positive prediction of \textsf{LTU} whereas a risk assessment of \textsf{medium} or \textsf{low} is equivalent to a negative prediction of \textsf{LTU}. 

To explore the effect of explanations on human-algorithm interaction, we ran a field experiment from October 2019 to June 2020. The assignment of treatment was randomized at the level of job centers.  Six centers were selected for the experiment, three for the treatment with 79 counselors and three for the control group with 77 counselors.  Within a job center, candidates were assigned counselors available at the time of registration.  After running the model, counselors were shown a risk assessment of \textsf{low}, \textsf{medium} or \textsf{high} and the raw probability score (risk score).  The treatment group of counselors was additionally shown \textsf{SHAP} which comprised a set of six features.  For a \textsf{high} (\textsf{low}) risk assessment, the top six features that increased (decreased) the probability of \textsf{LTU} were displayed; for a \textsf{medium} risk assessment, the top three features that, respectively, increased and decreased the probability of \textsf{LTU} were displayed. The counselors had the decisional authority and could either retain the algorithm's assessment or replace it with their own.  They were also asked to rate their confidence in the final assessment on a Likert scale of 1 to 5.  Data on the realized \textsf{LTU} outcomes of the candidates were collected in December 2021.  Further information on the empirical setting can be found in Appendices \ref{appendix a} and \ref{appendix b}.  

\subsection{Research Ethics and Social Impact}

Before launching the pilot, we obtained approval from the university's scientific council to run a study with human participants. All counselors taking part in the field experiment participated in a face-to-face information session.  A presentation on the the system they would be using was given and the counselors had the opportunity to ask the researchers questions.  Counselors were also provided the researchers’ e-mails and encouraged to get in touch with questions or concerns at any time.  All the counselors agreed to participate in the study.  After the end of the pilot, we conducted sessions presenting the results to representatives of all employment centers, which were recorded and made available to all PES (public employment service) counselors. 

The researchers did not have access to any personal data that could potentially identify, directly or indirectly, the PES users. All the PES identification numbers were pseudonymized by the PES.  All researchers with access to the data had training in personal data protection from the respective university's data protection office. 

\section{Methods}

\subsection{Identifying the Conflict} \label{4.1}

To extract a sparse causal representation effected by the algorithm in the human's perception, we regress the algorithm's \textsf{LTU} prediction $r$ on candidate features using \textsf{LASSO} logistic regression with ten-fold cross-validation for the control group.  The regression yields sparse linear models $p(r=i) = \mathbf{x}^\intercal{\boldsymbol{\beta}}_{i}$, where $p(r=i)$ is the probability of the prediction being $i$ (positive or negative), ${\boldsymbol{\beta}}_{i}$ the set of coefficients associated with prediction $i$, and $\mathbf{x}$ the sparse set of features significant for driving variances in the algorithm's \textsf{LTU} predictions.  

\begin{table}[h]
\caption{LASSO regression coefficients for the representation attributed to the algorithm}
\begin{center}
\begin{tabular}{rlrr}
 \hline 
 \hline \\[-1.8ex] 
 & \textsf{LTU} prediction & \textsf{non-LTU}  & \textsf{LTU}\\  
  \\[-1.8ex] \hline \\[-1.8ex] 
  1 & \textsf{reason = contract ended} & 0.304 & $-$0.304 \\ 
  2 & \textsf{reason = mutual agreement} & $-$0.347 & 0.347 \\
  3 & \textsf{education = college} & $-$0.127 & 0.127  \\
  4 & \textsf{social integration subsidy = true} & $-$1.163 & 1.163 \\
  5 & \textsf{age} & $-$0.133 & 0.133 \\ 
  6 & \textsf{number of registrations} & 0.037 & $-$0.037 \\ 
  7 & \textsf{number of subsidy suspensions} & 0.012 & $-$0.012 \\ 
  8 & \textsf{age group : $>$ 56} & $-$0.518 & 0.518 \\ 
  9 & \textsf{unemployment length} & $-$0.023 & 0.023 \\ 
  \\[-1.8ex]\hline 
  \hline \\[-1.8ex] 
\end{tabular}
\end{center}
\label{table1}
\end{table}

Of the nine features, \textsf{age}, \textsf{number of registrations}, \textsf{number of subsidy suspensions}, and \textsf{unemployment length} are numeric variables.  The rest are dummy variables derived from categorical variables.  The representation attributed to the algorithm has the following possible semantics: if the unemployed candidate left her previous employment by mutual agreement, is college-educated, receives social integration subsidy, is in the age group above 56, is older, has had a fewer number of registrations, has had a fewer number of subsidy suspensions, and/or has been unemployed for longer, then she is more likely to be (judged by the algorithm to be) in long-term unemployment.  However, if the unemployed candidate left her previous employment because the contract ended, is younger, has had a greater number of registrations, has had a greater number of subsidy suspensions, and/or has been unemployed for shorter, then she is less likely to be (judged by the algorithm to be) in long-term unemployment. 
\begin{align*}
&R_{a}: \\
&\{\textsf{reason} = \textsf{mutual agreement}\} \lor \{\textsf{education}=\textsf{college}\} \lor \\
&\{\textsf{social integration subsidy} = \textsf{true}\} \lor 
\{\textsf{age group} : \textsf{$>$ 56}\} \lor \{\textsf{age +}\} \lor \\
&\{\textsf{number of registrations -}\} \lor 
\{\textsf{number of subsidy suspensions -}\} \lor \\
&\{\textsf{unemployment length +}\} \mapsto \textsf{LTU}; \\
&\{\textsf{reason} = \textsf{contract ended}\} \lor \{\textsf{age -}\} \lor \{\textsf{number of registrations +}\} \lor \\
&\{\textsf{number of subsidy suspensions +}\} \lor \{\textsf{unemployment length -}\} \mapsto \textsf{non-LTU}.\\
\end{align*} 

The counselor either retains or adjusts the algorithm's risk assessment.  We construct a variable $a$, with actions $a = -1$ (adjusting the algorithm's risk assessment downward), $a = 0$ (retaining the algorithm's risk assessment as it is), and $a = 1$ (adjusting the algorithm's risk assessment upward).  To identify features where the counselor have a strong prior belief, we regress $a$ on candidate features using \textsf{LASSO} multinomial regression with ten-fold cross-validation for the control group.  This yields sparse linear models $p(a=j) = \mathbf{w}^\intercal{\boldsymbol{\alpha}}_{j}$, where $p(a=j)$ is the probability of action $j$, ${\boldsymbol{\alpha}}_{j}$ the set of coefficients associated with action $j$, and $\mathbf{w}$ the sparse set of features likely, ceteris paribus, to induce the counselors to take a particular action.  The regression is run over the counselors as an aggregate so the prior belief is assumed to be held collectively.  Both the sparsity and the linearity dovetail with the inductive bias for human mental models, which tend to be sparse and linear \cite{lombrozo2007simplicity}    

\begin{table}[h] 
\caption{LASSO regression coefficients for the human prior belief}
\begin{center}
\begin{tabular}{@{\extracolsep{5pt}}rlrrr}
 \hline 
 \hline \\[-1.8ex] 
 & action & down & same & up \\ 
  \\[-1.8ex] \hline \\[-1.8ex] 
  1 & \textsf{reason = was student} & 0.099 & $-$0.061 & $-$0.037 \\ 
  2 & \textsf{nationality = non-EU/EEA} & $-$0.007 & $-$0.027 & 0.034 \\
  3 & \textsf{education = college} & 0.464 & $-$0.075 & $-$0.389 \\
  4 & \textsf{age} & $-$0.001 & 0.004 & $-$0.004 \\
  5 & \textsf{desired job = scientific research} & 0.012 & $-$0.005 & $-$0.008 \\
  6 & \textsf{personal employment plan = true} & 0.196 & $-$0.131 & $-$0.065 \\
  7 & \textsf{prior personal employment plan = true} & $-$0.022 & 0.034 & $-$0.012 \\
  8 & \textsf{number of interventions in job training} & 0.061 & $-$0.034 & $-$0.027 \\ 
  9 & \textsf{was LTU = true} & 0.052 & 0.064 & $-$0.116 \\ 
  \\[-1.8ex]\hline 
  \hline \\[-1.8ex] 
\end{tabular}
\end{center}
\label{table2}
\end{table}

Except for \textsf{age} and \textsf{number of interventions in job training}, which are numeric variables, all are dummy variables derived from categorical variables.  The human prior belief has the following semantics: if the unemployed candidate left her previous employment because she was a student, is college-educated, desires to find a job in scientific research, has a personal employment plan, and/or has had a greater number of interventions in job training, then she is less likely to be in long-term unemployment.  However, if the candidate does not have EU/EEA nationality and/or has had a fewer number of interventions in job training, then she is more likely to be in long-term unemployment.  The features \textsf{age}, \textsf{prior personal employment plan = true}, and \textsf{was LTU = true} have positive coefficients for $a = 0$ (retaining the algorithm's prediction).  Thus there is not any strongly conflicting prior belief associated with them, although \textsf{was LTU = true} is slightly ambiguous as it also has a positive coefficient for downward adjustment.   

\begin{align*}
&R_{h}: \\
&\{\textsf{reason} = \textsf{was student}\} \lor \{\textsf{education}=\textsf{college}\} \lor \\ 
&\{\textsf{desired job} = \textsf{scientific research}\} \lor 
\{\textsf{personal employment plan} = \textsf{true}\} \lor \\
&\{\textsf{number of interventions in job training + }\} 
\mapsto \textsf{non-LTU}; \\
& \{\textsf{nationality} = \textsf{non-EU/EEA}\} \lor \\ 
&\{\textsf{number of interventions in job training -}\}  \mapsto \textsf{LTU}.
\end{align*}

Comparing $R_{a}$ and $R_{h}$, we identify \textsf{college education} as the feature where the human prior belief conflicts with the causal representation effected by the algorithm as a statistical regularity.  Whereas the algorithm effects a sparse causal representation college education to a higher risk of long-term unemployment, the human prior belief seems to be of the opposite view --- that college education should lead to a lower risk of long-term unemployment.    

\subsection{Identifying the Effects of Explanations on Decision Quality and Confidence} \label{4.2}

We first partition the experimental data into instances for which the algorithm gives a positive prediction of \textsf{LTU} (a risk assessment of \textsf{high}) and instances for which the algorithm gives a negative prediction of \textsf{LTU} (a risk assessment of \textsf{medium} or \textsf{low}).  This is done for two reasons.  Firstly, a conflict is specified only for a given algorithmic prediction.  Secondly, the covariates and dependent variables in our regressions have different distributions for the two subsets of observations.      

We construct a dummy variable $\textsf{Conflict}$ to indicate the presence or absence of features where $R_{a}$ and $R_{h}$ conflict with each other in the explanation (shown to the treatment but not the control group).  We estimate the following equation as our first model using logistic regression: 
\begin{align*} \label{eq1}
\textsf{Decision Quality} &= \gamma_0 + \gamma_1 \textsf{Exposed} + \gamma_2 \textsf{Conflict} + \gamma_3 \textsf{Exposed} \times \textsf{Conflict} \\
&+ \gamma_4\textsf{Risk Score} + \textsf{time fixed effects} + \varepsilon.  
\tag{1}     
\end{align*}
where $\textsf{Decision Quality}$ is the accuracy of the final assessment and $\textsf{Exposed}$ indicates treatment status.  We use $\textsf{Risk Score}$, which is native to \textsf{XGBoost}, as a control variable that stratifies the data instances into bins of predictions of equal difficulty.  The sign and statistical significance of $\gamma_3$ tell us whether showing features on which the counselors hold a conflicting prior belief as part of the explanation affects the decision quality. 

A second model, similarly estimated using logistic regression, explores the heterogeneity of the mechanism by adding $\textsf{Adjustment}$ as an interaction variable, where $\textsf{Adjustment}$ is a dummy variable indicating whether the counselor has adjusted the algorithm's \textsf{LTU} prediction: 
\begin{align*} \label{eq2}
\textsf{Decision Quality} &= \delta_0 + \delta_1 \textsf{Exposed} + \delta_2 \textsf{Conflict} + \delta_3 \textsf{Adjustment} \\
&+ \delta_4 \textsf{Exposed} \times \textsf{Conflict} + \delta_5 \textsf{Exposed} \times \textsf{Adjustment} \\ 
&+ \delta_6 \textsf{Conflict} \times \textsf{Adjustment} + \delta_7 \textsf{Exposed} \times \textsf{Conflict} \times \textsf{Adjustment} \\ &+ \delta_8\textsf{Risk Score} + \textsf{time fixed effects} + \eta.  \tag{2}    
\end{align*}
The sign and statistical significance of $\delta_4$ ($\delta_7$) tell us whether showing features on which the counselors hold a conflicting prior belief as part of the explanation affects the decision quality, when the counselor retains (adjusts) the algorithm's prediction. 

In a third model we examine the impact of explanations on counselors' confidence in the final assessment using linear regression:

\begin{align*} \label{eq3}
\textsf{Confidence} &= \zeta_0 + \zeta_1 \textsf{Exposed} + \zeta_2 \textsf{Conflict} + \zeta_3 \textsf{Adjustment} \\
&+ \zeta_4 \textsf{Exposed} \times \textsf{Conflict} + \zeta_5 \textsf{Exposed} \times \textsf{Adjustment} \\ 
&+ \zeta_6 \textsf{Conflict} \times \textsf{Adjustment} + \zeta_7 \textsf{Exposed} \times \textsf{Conflict} \times \textsf{Adjustment} \\ &+ \zeta_8\textsf{Risk Score} + \textsf{time fixed effects} + \xi. \tag{3} 
\end{align*} 

The sign and statistical significance of $\zeta_4$ ($\zeta_7$) tell us whether showing features on which the counselors hold a conflicting prior belief as part of the explanation increases or decreases confidence, when the counselor retains (adjusts) the algorithm's prediction.  

Models \ref{eq1} and \ref{eq2} are estimated as linear probability models and model \ref{eq3} is estimated as a linear regression model.

\section{Results}

For model \ref{eq1}, $\textsf{Conflict}$ is equal to $1$ when \textsf{college education} is displayed as part of the explanation (or would have been displayed to the control group as part of the explanation had their treatment condition been different).  Only the subset of instances for which the algorithm gives a positive prediction of \textsf{LTU} is included.  The regression results show that a positive prediction with a higher $\textsf{Risk Score}$ is more likely to be correct.  The coefficient for $\textsf{Exposed} \times \textsf{Conflict}$ is negative and statistically significant.  This means that displaying \textsf{college education} as part of the explanation degrades the quality of decision-making. 

\begin{table}[h]
\caption{Regression results}
\begin{tabular}{@{\extracolsep{5pt}}lccc} 
\\[-1.8ex]\hline 
\hline \\[-1.8ex] 
 & \multicolumn{3}{c}{\textit{Dependent variable:}} \\ 
\cline{2-4} 
\\[-1.8ex] & Decision Quality & Decision Quality & Confidence \\ 
\\[-1.8ex] & LPM & LPM & Linear \\ 
\\[-1.8ex] & (1) & (2) & (3)\\ 
\hline \\[-1.8ex] 
 Exposed &  $-$0.035* & $-$0.024 & -0.006 \\ 
    & (0.015)   & (0.016) & (0.018) \\ 
 Exposed $\times$ Conflict & $-$0.104*** & $-$0.049 & 0.171*** \\
        &  (0.030)   & (0.038) & (0.041) \\
 Exposed $\times$ Conflict $\times$ Adjustment & & $-$0.166** & $-$0.331***\\
        & &  (0.058)  & (0.063)  \\
 Conflict & 0.031 & $-$0.009 & $-$0.100** \\
        &  (0.023)   & (0.028) & (0.031) \\
 Adjustment & & $-$0.433*** & $-$0.191*** \\
        &  & (0.024) & (0.027) \\
 Exposed $\times$ Adjustment & & 0.047 & 0.108** \\
        &  & (0.031) & (0.034) \\       
 Conflict $\times$ Adjustment & & 0.277*** & 0.349*** \\
        &  & (0.044) & (0.048) \\
 Risk Score &  0.834*** & 0.460*** & 0.164\\
       & (0.085) & (0.081) & (0.089) \\ 
 2020Q1 &  0.084*** &  0.082***  & $-$0.030\\
        &  (0.016)   &  (0.015) & (0.016) \\   
 2020Q2 &  0.076*** & 0.084*** & $-$0.199*** \\
     &      (0.016) & (0.015) & (0.017) \\
\hline \\[-1.8ex] 
Observations & 5,728 & 5,728 & 5,728 \\ 
R$^{2}$ & 0.029 & 0.1446 & 0.048\\ 
Adjusted R$^{2}$ & 0.028 & 0.1431 & 0.046\\ 
\hline 
\hline \\[-1.8ex] 
\textit{Note:}  & \multicolumn{3}{r}{$^{*}$p$<$0.05; $^{**}$p$<$0.01; $^{***}$p$<$0.001} \\ 
 & \multicolumn{3}{r}{} \\ 
\end{tabular} 
\end{table}

Regression results for model \ref{eq2} show that there is heterogeneity in the effects of explanations on decision quality with the larger part of the decrease coming from when counselors adjust the algorithm's positive prediction of \textsf{LTU}, as indicated by the negative and statistically significant coefficient for $\textsf{Exposed} \times \textsf{Conflict} \times \textsf{Adjustment}$. 
 
Finally, regression results for model \ref{eq3} show that displaying \textsf{college education} as part of the explanation has polarizing effects on confidence.  It reduces confusion and increases confidence when counselors retain the algorithm's prediction, as indicated by the positive and statistically significant coefficient for $\textsf{Exposed} \times \textsf{Conflict}$.  On the other hand it draws attention to the conflict and decreases confidence when counselors adjust the algorithm's prediction, as indicated by the negative and statistically significant coefficient for $\textsf{Exposed} \times \textsf{Conflict} \times \textsf{Adjustment}$.              

\section{Discussion} \label{discussion}

Conflict between epistemic standpoints defines the kind of rationality that enables actions.  If there is no \textit{epistemic conflict}, agency --- the ability to perform a difference-making action as it pertains to one's goal --- would not be realized.  

In our field experiment, the conflict between the human's prior belief that college education is negatively associated with long-term unemployment and the obverse representation attributed to the algorithm leads to actions that worsen the quality of decision-making for the time period during which the pilot was run.  An alternative scenario, however, can be conceived where the human's prior beliefs encode useful information about the world that the algorithm is not privy to.  In such a scenario, epistemic conflict can improve the quality of decision-making.  Nevertheless it is not possible for an organization to determine ex-ante whether epistemic conflict would degrade or enhance the quality of decision-making.

Explanations perform an epistemic or communicative function by rendering complex representations human-interpretable.  This however does not resolve the conflict of representations and may in fact --- because explanations direct attention to cogent representations with explicit semantics --- exacerbate it.  As can be seen with positive predictions of \textsf{LTU} in our field experiment, showing \textsf{college education} as part of the explanation degrades the quality of decision-making.

Bare communication of representations, therefore, is not sufficient.  Such communication is oddly solipsistic in that each representation is committed to its own epistemic standpoint.  What should drive human-algorithm interaction is not the mere fact of epistemic conflict but its \textit{whys}, just as human beings do not just insist on their differences but act to understand and bridge them.  A more extensive communicative rationality is needed to enable actions that would improve the quality of decision-making.

We believe there are four desiderata for such communicative rationality.  The first desideratum is \textit{understanding}.  The human should understand the algorithm's epistemic standpoint.  In our empirical context, this means being imparted information about model training as well as possible reasons for certain representations (e.g. \textit{why} college education is associated with higher risk of long-term unemployment).  The second desideratum is \textit{reciprocity}.  Human-AI interaction tends to be unidirectional. 
Increasing the algorithm's understanding of the human's epistemic standpoint and prior beliefs can improve communication.  The third desideratum is \textit{negotiability}.  Reciprocal understanding facilitates negotiation where arguments can be developed, evidences arrayed, biases identified, and confidences gauged.  The last desideratum is a \textit{shared reality}.  In our empirical context, the counselors do not receive feedback on the accuracy of their judgment.  Convergence to a shared reality is possible if the counselors are made aware of where each of the two parties has erred.

\newpage


\bibliographystyle{splncs04}
\bibliography{bibliography}

\begin{thebibliography}{10}
\providecommand{\url}[1]{\texttt{#1}}
\providecommand{\urlprefix}{URL }
\providecommand{\doi}[1]{https://doi.org/#1}

\bibitem{bundorf2019humans}
Bundorf, K., Polyakova, M., Tai-Seale, M.: How do humans interact with
  algorithms? experimental evidence from health insurance. Tech. rep., National
  Bureau of Economic Research (2019)

\bibitem{colin2022cannot}
Colin, J., Fel, T., Cad{\`e}ne, R., Serre, T.: What i cannot predict, i do not
  understand: A human-centered evaluation framework for explainability methods.
  Advances in Neural Information Processing Systems  \textbf{35},  2832--2845
  (2022)

\bibitem{cyert1963behavioral}
Cyert, R.M., March, J.G., et~al.: A behavioral theory of the firm, vol.~2.
  Englewood Cliffs, NJ (1963)

\bibitem{delanda2021materialist}
DeLanda, M.: Materialist Phenomenology: A Philosophy of Perception. Bloomsbury
  Publishing (2021)

\bibitem{dietvorst2015algorithm}
Dietvorst, B.J., Simmons, J.P., Massey, C.: Algorithm aversion: People
  erroneously avoid algorithms after seeing them err. Journal of Experimental
  Psychology: General  \textbf{144}(1), ~114 (2015)

\bibitem{dietvorst2018overcoming}
Dietvorst, B.J., Simmons, J.P., Massey, C.: Overcoming algorithm aversion:
  People will use imperfect algorithms if they can (even slightly) modify them.
  Management Science  \textbf{64}(3),  1155--1170 (2018)

\bibitem{fiske1982structural}
Fiske, S.T., Kenny, D.A., Taylor, S.E.: Structural models for the mediation of
  salience effects on attribution. Journal of Experimental Social Psychology
  \textbf{18}(2),  105--127 (1982)

\bibitem{fugener2021will}
F{\"u}gener, A., Grahl, J., Gupta, A., Ketter, W.: Will humans-in-the-loop
  become borgs? merits and pitfalls of working with ai. Management Information
  Systems Quarterly (MISQ)-Vol  \textbf{45} (2021)

\bibitem{fugener2022cognitive}
F{\"u}gener, A., Grahl, J., Gupta, A., Ketter, W.: Cognitive challenges in
  human--artificial intelligence collaboration: investigating the path toward
  productive delegation. Information Systems Research  \textbf{33}(2),
  678--696 (2022)

\bibitem{galhotra2021explaining}
Galhotra, S., Pradhan, R., Salimi, B.: Explaining black-box algorithms using
  probabilistic contrastive counterfactuals. In: Proceedings of the 2021
  International Conference on Management of Data. pp. 577--590 (2021)

\bibitem{gao2021human}
Gao, R., Saar-Tsechansky, M., De-Arteaga, M., Han, L., Lee, M.K., Lease, M.:
  Human-ai collaboration with bandit feedback. arXiv preprint arXiv:2105.10614
  (2021)

\bibitem{gigerenzer1996reasoning}
Gigerenzer, G., Goldstein, D.G.: Reasoning the fast and frugal way: models of
  bounded rationality. Psychological review  \textbf{103}(4), ~650 (1996)

\bibitem{Gigerenzer:1999p256}
Gigerenzer, G., Todd, P., Group, A.: Simple heuristics that make us smart
  (1999)

\bibitem{glikson2020human}
Glikson, E., Woolley, A.W.: Human trust in artificial intelligence: Review of
  empirical research. Academy of Management Annals  \textbf{14}(2),  627--660
  (2020)

\bibitem{guidotti2021evaluating}
Guidotti, R.: Evaluating local explanation methods on ground truth. Artificial
  Intelligence  \textbf{291},  103428 (2021)

\bibitem{han2022explanation}
Han, T., Srinivas, S., Lakkaraju, H.: Which explanation should i choose? a
  function approximation perspective to characterizing post hoc explanations.
  arXiv preprint arXiv:2206.01254  (2022)

\bibitem{jacovi2021formalizing}
Jacovi, A., Marasovi{\'c}, A., Miller, T., Goldberg, Y.: Formalizing trust in
  artificial intelligence: Prerequisites, causes and goals of human trust in
  ai. In: Proceedings of the 2021 ACM Conference on Fairness, Accountability,
  and Transparency. pp. 624--635 (2021)

\bibitem{kawaguchi2021will}
Kawaguchi, K.: When will workers follow an algorithm? a field experiment with a
  retail business. Management Science  \textbf{67}(3),  1670--1695 (2021)

\bibitem{lagnado_2021}
Lagnado, D.A.: Explaining the Evidence: How the Mind Investigates the World.
  Cambridge University Press (2021). \doi{10.1017/9780511794520}

\bibitem{lebovitz2022engage}
Lebovitz, S., Lifshitz-Assaf, H., Levina, N.: To engage or not to engage with
  ai for critical judgments: How professionals deal with opacity when using ai
  for medical diagnosis. Organization Science  (2022)

\bibitem{lombrozo2007simplicity}
Lombrozo, T.: Simplicity and probability in causal explanation. Cognitive
  psychology  \textbf{55}(3),  232--257 (2007)

\bibitem{lu2008bayesian}
Lu, H., Yuille, A.L., Liljeholm, M., Cheng, P.W., Holyoak, K.J.: Bayesian
  generic priors for causal learning. Psychological review  \textbf{115}(4),
  ~955 (2008)

\bibitem{lundberg2017unified}
Lundberg, S.M., Lee, S.I.: A unified approach to interpreting model
  predictions. In: Advances in neural information processing systems. pp.
  4765--4774 (2017)

\bibitem{nauta2022anecdotal}
Nauta, M., Trienes, J., Pathak, S., Nguyen, E., Peters, M., Schmitt, Y.,
  Schl{\"o}tterer, J., van Keulen, M., Seifert, C.: From anecdotal evidence to
  quantitative evaluation methods: A systematic review on evaluating
  explainable ai. arXiv preprint arXiv:2201.08164  (2022)

\bibitem{pearl2018book}
Pearl, J., Mackenzie, D.: The book of why: the new science of cause and effect.
  Basic books (2018)

\bibitem{ribeiro2016should}
Ribeiro, M.T., Singh, S., Guestrin, C.: " why should i trust you?" explaining
  the predictions of any classifier. In: Proceedings of the 22nd ACM SIGKDD
  international conference on knowledge discovery and data mining. pp.
  1135--1144 (2016)

\bibitem{schwab2019cxplain}
Schwab, P., Karlen, W.: Cxplain: Causal explanations for model interpretation
  under uncertainty. Advances in Neural Information Processing Systems
  \textbf{32} (2019)

\bibitem{simon2013administrative}
Simon, H.A.: Administrative behavior. Simon and Schuster (2013)

\bibitem{slack2021reliable}
Slack, D.Z., Hilgard, S., Singh, S., Lakkaraju, H.: Reliable post hoc
  explanations: Modeling uncertainty in explainability. In: Beygelzimer, A.,
  Dauphin, Y., Liang, P., Vaughan, J.W. (eds.) Advances in Neural Information
  Processing Systems (2021), \url{https://openreview.net/forum?id=rqfq0CYIekd}

\bibitem{sun2022predicting}
Sun, J., Zhang, D.J., Hu, H., Van~Mieghem, J.A.: Predicting human discretion to
  adjust algorithmic prescription: A large-scale field experiment in warehouse
  operations. Management Science  \textbf{68}(2),  846--865 (2022)

\bibitem{taylor1979generalizability}
Taylor, S.E., Crocker, J., Fiske, S.T., Sprinzen, M., Winkler, J.D.: The
  generalizability of salience effects. Journal of Personality and Social
  Psychology  \textbf{37}(3), ~357 (1979)

\bibitem{taylor1975point}
Taylor, S.E., Fiske, S.T.: Point of view and perceptions of causality. Journal
  of Personality and Social Psychology  \textbf{32}(3), ~439 (1975)

\bibitem{taylor1978salience}
Taylor, S.E., Fiske, S.T.: Salience, attention, and attribution: Top of the
  head phenomena. In: Advances in experimental social psychology, vol.~11, pp.
  249--288. Elsevier (1978)

\bibitem{ullman2018does}
Ullman, D., Malle, B.F.: What does it mean to trust a robot? steps toward a
  multidimensional measure of trust. In: Companion of the 2018 acm/ieee
  international conference on human-robot interaction. pp. 263--264 (2018)

\bibitem{wan2022explainability}
Wan, C., Belo, R., Zejnilovic, L.: Explainability's gain is optimality's loss?
  how explanations bias decision-making. In: Proceedings of the 2022 AAAI/ACM
  Conference on AI, Ethics, and Society. pp. 778--787 (2022)

\bibitem{woodward2021causation}
Woodward, J.: Causation with a human face: Normative theory and descriptive
  psychology. Oxford University Press (2021)

\bibitem{xiang2022collaborative}
Xiang, Y., V{\'e}lez, N., Gershman, S.J.: Collaborative decision making is
  grounded in representations of other people's competence and effort  (2022)

\end{thebibliography}

\appendix

\newpage
\section{Empirical Setting} \label{appendix a}

\begin{table}[h]
\caption{Treatment assignment to job centers}
\label{job_centers}
\begin{center}
\begin{tabular}{@{}c|cc|cc|c@{}}
\hline 
& \multicolumn{2}{c|}{pre-pilot} & \multicolumn{2}{c|}{pilot} & \multicolumn{1}{c}{} \\ \hline
job center & registrations & appointments/mo. & registrations & appointments/mo. & \multicolumn{1}{c}{treatment} \\ \hline
1 & 11958 & 213 & 13139 & 169 & 0 \\
2 & 9406 & 191 & 10263 & 160 & 1 \\
3 & 3396 & 99 & 3743 & 72 & 0 \\
4 & 5717 & 110 & 6022 & 88 & 1 \\
5 & 3889 & 78 & 4379 & 69 & 0 \\
6 & 7016 & 135 & 7336 & 100 & 1 \\ \hline
\end{tabular}
\end{center}
\end{table}

\newpage

\section{User Interface} \label{appendix b}

\begin{figure}[h]
\centering
    \includegraphics[scale=0.40]{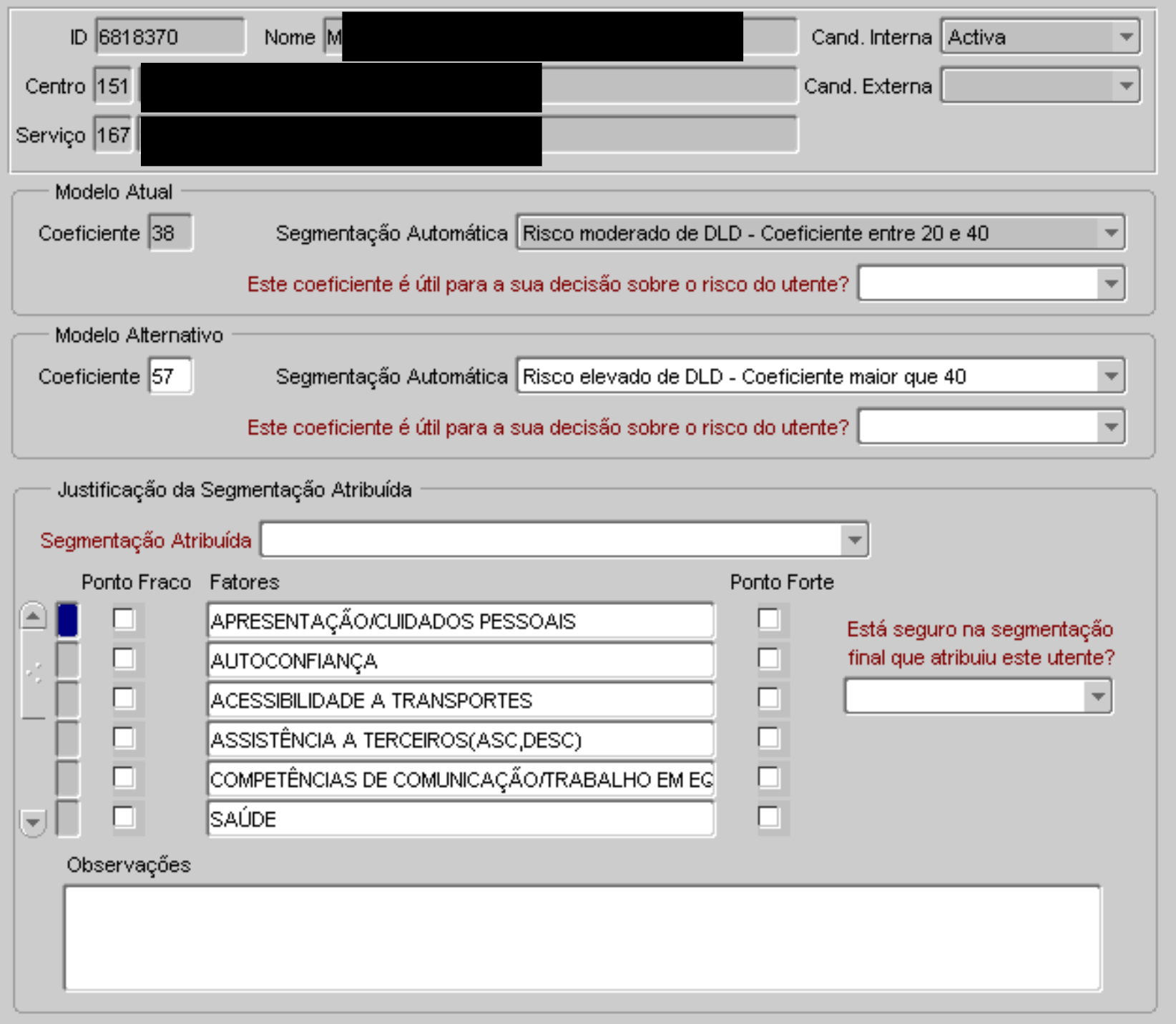}
     \label{fig:ui-control}
     \caption{User Interface for the Control Group.  In the ``Modelo Atual'' panel the risk score and the risk assessment are shown.  In the ``Justificação da Segemntação Atribuída'' panel the counselor has to select her own risk assessment on the top and her confidence level on the right.}
\end{figure}

\newpage

\begin{figure}[H]
\centering
    \includegraphics[scale=0.40]{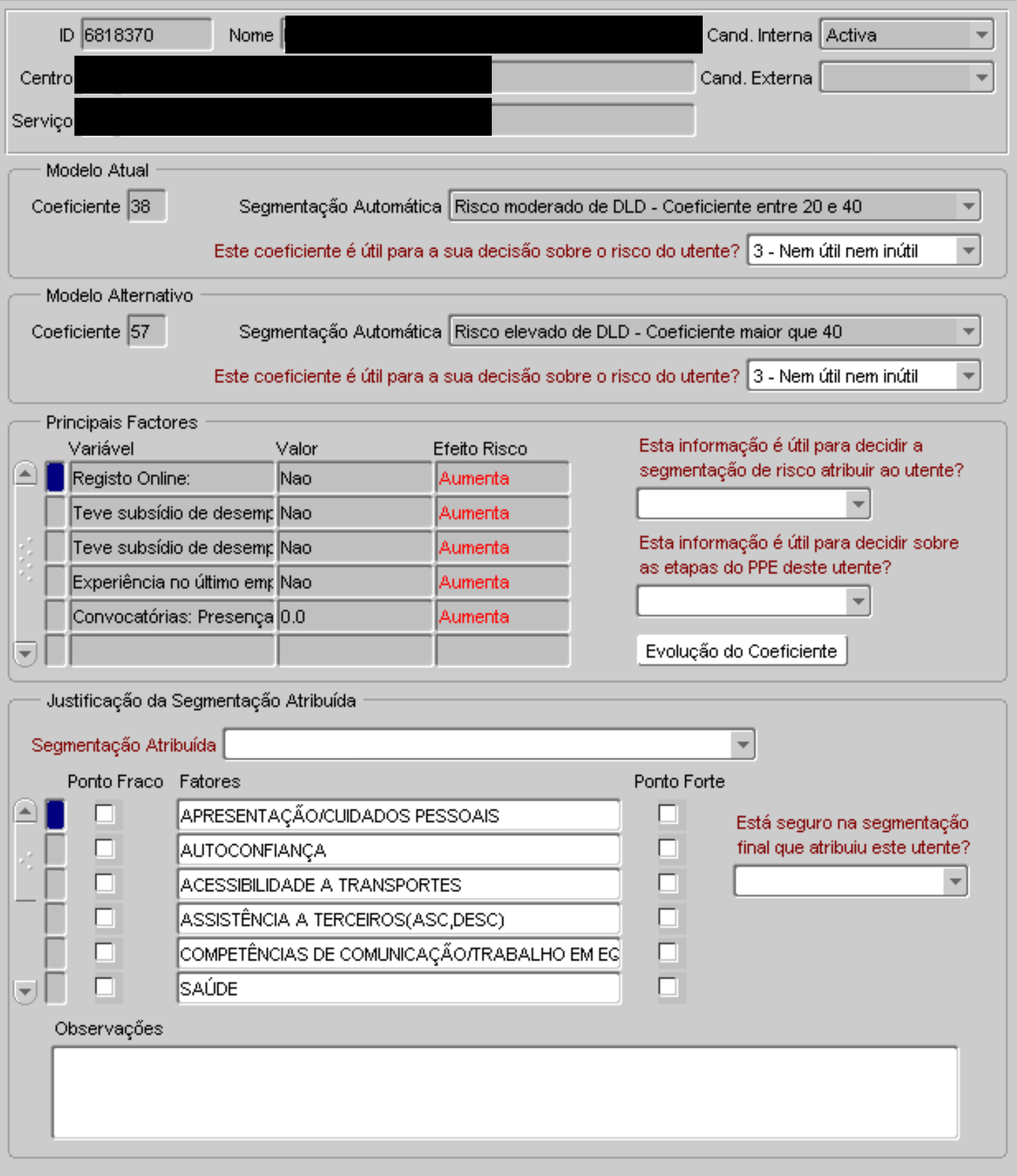}
     \label{fig:ui-control}
     \caption{User Interface for the Treatment Group.  \textsf{SHAP} are shown in the ``Principais Factores'' panel with their respective effect on the risk of \textsf{LTU}.}
\end{figure}

\end{document}